# A Doorway mechanism for Electron Attachment Induced DNA Strand Break

*Jishnu Narayanan S J, Divya Tripathi, and Achintya Kumar Dutta\**

*Department of Chemistry, Indian Institute of Technology Bombay, Powai, Mumbai 400076.*

**Abstract**

We report a new doorway mechanism for the dissociative electron attachment to genetic materials. The dipole-bound state of the nucleotide anion acts as the doorway for electron capture in the genetic material. The electron gets subsequently transferred to a dissociative σ* type anionic state localized on a sugar-phosphate or a sugar-nucleobase bond, leading to their cleavage. The electron transfer is mediated by the mixing of electronic and nuclear degrees of freedom. The cleavage rate of the sugar-phosphate bond predicted by this new mechanism is higher than that of the sugar-nucleobase bond breaking, and both processes are considerably slower than the formation of a stable valence-bound anion. The new mechanism explains the relative rates of electron attachment induced bond cleavages in genetic materials.

*achintya@chem.iitb.ac.in

Radiation therapy is one of the most popular means of treating cancer. It involves exposing the tumor cells to ionizing radiation such as X-, γ-, or β-radiations, which leads to DNA damage and ultimately to cell death.[1] The high-energy photons interact with DNA and its aqueous environment in the cell. It results in the formation of reactive species such as free radicals, ions, secondary electrons, which are potentially harmful to DNA. It is widely accepted that the contribution of the secondary products is higher than direct radiation-DNA interaction in causing damage.[2,3] Among them, the role of secondary electrons in radiation damage to genetic material has attracted considerable attention recently.[3–7] Once formed, the secondary electrons rapidly lose energy because of inelastic collisions with the solvent molecules. These Low energy electrons (LEE) with energies <20eV are then taken up by the DNA, which leads to single-strand breaks (SSB), double-strand breaks (DSB), or even clustered DNA damages.[8–21] Even though it is widely accepted that LEEs cause DNA damage, the exact mechanism of this process is still not precisely understood. A deep insight into this process is vital for enhancing the efficiency of radiation therapy.

Sanche and coworkers' pioneering work on the role of LEEs in DNA SSBs and DSBs ignited a series of exciting experimental and theoretical investigations regarding the process.[9–13,22–30] They reported that the LEEs with energies well below the ionization thresholds of DNA could induce strand breaks.[8] Following this report, numerous attempts[23–27] have been made to understand the mechanism of the dissociative electron attachment (DEA). A majority of them proposed the existence of short-lived shape resonance states located on nucleobase that lead to SSB.[23] Nucleobase localized shape resonances lie in the energy range 0.1-2 eV and are accessible by the LEEs. The electron from the π*-MO shape resonance state subsequently gets transferred to sugar-phosphate (C-O)/ sugar-nucleobase (C-N) σ*-MO, which results in the rupture of the C-O/C-N bond.[27] Sanche and coworkers concluded that this short-lived transient negative ion (TNI) could not result in the C-N bond rupture.[31] It is the long-lived core excited resonance that causes the cleavage of the C-N bond. The sugar-phosphate bond rupture can also happen upon the direct electron attachment to the phosphate group. Kumar *et al.* showed the existence of dissociative σ* levels accessible to electrons within the energy range >2eV from the π* orbital of P=O bond.[32]

Many of these DNA bases have a sizeable dipole moment which can support dipole-bound anionic states in the gas phase.[33] The dipole-bound state can act as a doorway to the valence-bound state of the nucleobases.[34] The two states are clearly distinguishable because the extra electron density is located away from the nuclear framework in a dipole-bound state. Whereas in the valence-bound anion, the extra electron density is localized on the nucleobase (Figure 1). We have recently shown that the doorway mechanism is a generalized path to the formation of stable nucleobase anion,[34] and they even exist in the condensed phase, where the solvent-bound states act as the doorway.[35–37] The doorway mechanism is traditionally used to explain the formation of bound anionic states. In this work, we have explored the possibility of a doorway mechanism for the LEE-induced DEA process in genetic materials. Previous studies have shown that purine nucleobases have the highest vertical electron affinity among nucleobases.[33,38] Between cytosine and thymine, the vertically bound anion of the former is more stable than that of the latter.[38] Therefore, we have considered cytosine nucleotide as a simple model system. Figure 1 presents the dipole- and valence-bound states of 3′-deoxycytidine monophosphate (3′-dCMPH). The difference in the location of the additional electron density leads to a striking contrast between the geometry of the two anionic states. Previous studies have shown that the geometry of the neutral molecules mostly remains unperturbed due to the formation of the dipole-bound anion.[33,34] Hence, we have assumed the structure of the neutral species same as the dipole-bound anion's geometry in this study. The valence-bound geometry, on the contrary, is visibly different from the neutral structure with an RMSD of 0.711 Å. The elongation of the bonds on the cytosine ring further confirms that the extra electron occupies the π*-antibonding molecular orbital in the valence-bound 3′-dCMPH anion. The dipole-bound state is vertically bound with vertical detachment energy (VDE) of 0.106 eV. On the other hand, the valence bound state is only bound adiabatically with a VDE of 1.089 eV.

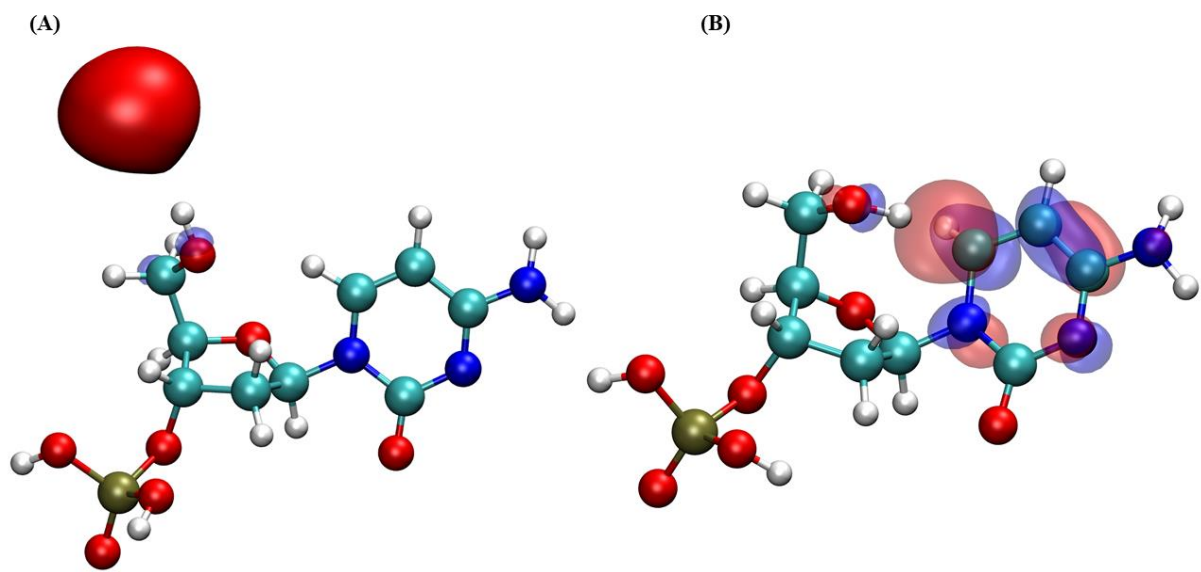

**Figure 1. The molecular orbital corresponding to the dominant transition in the EA-EOM-DLPNO-CCSD for (A) dipole-bound and B) valence-bound deoxycytidine-3′-monophosphate anion.**

The formation of the dipole-bound doorway anionic state of 3′-dCMPH can lead to two distinct events. The electron may get transferred to the π*-antibonding molecular orbital localized on the nucleobase leading to the formation of a valence-bound state. Alternatively, the electron may also get transferred to the σ*-MO of C-O or C-N, resulting in the dissociation of the corresponding bond. The transition of the dipole-bound electron to the π*-antibonding molecular orbital is very feebly optically allowed. However, the process may be assisted by the vibrational degrees of freedom of the molecule. Modeling a potential energy surface to describe this process accurately is tricky due to the multidimensional nature of the problem. One can work around this issue by reducing its dimensionality. We have constructed a one-dimensional adiabatic potential energy surface along a linear trajectory between the dipole-bound and valence-bound states (See Figure 2). The geometrical parameters (bond length, bond angle, etc.) are varied linearly (according to Equation (1)) to obtain the intermediate geometries along the trajectory. The transition between the two electronic states leads to an avoided crossing in the adiabatic potential energy surface, where the Born-Oppenheimer approximation breaks down.[39] We have modeled a diabatic surface by choosing the dipole-bound and valence-bound anionic states as the basis and calculated the coupling element using a two-state model Hamiltonian. The diagonal terms are approximated using Morse potential. The value of the coupling element between dipole- and valence-bound potential energy curve (PEC) for 3′-dCMPH is 17.86 meV, which indicates the presence of weak coupling between the two states. One can use the Marcus theory to get an estimate of the rate of electron transfer from dipole-bound state to valence-bound state,[40] which in the present case has been found to be $5.88 \times 10^7 \, \text{s}^{-1}$ at $T = 293$ K. It indicates that the dipole-bound state can act as a doorway to the valence-bound state in nucleotides.

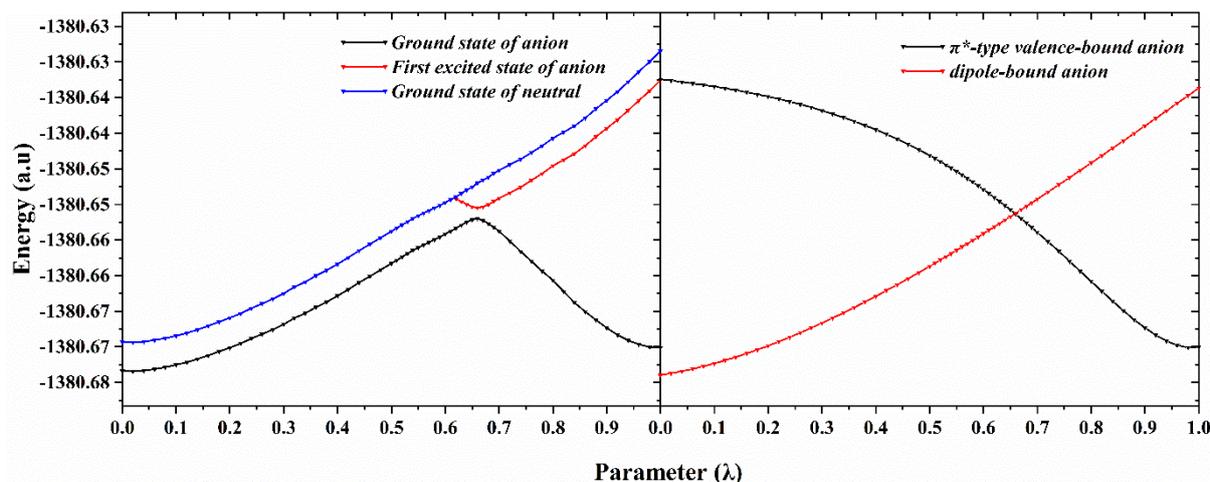

**Figure 2. a) Adiabatic and b) diabatic PEC corresponding to the dipole-bound to valence-bound transition for 3′-dCMPH. $\lambda$ is the parameter varied from 0 to 1 to obtain the intermediate geometries.**

Now, the formation of the dipole-bound anion can also be followed by C-O and/or C-N bond cleavage. Based on experimental studies, LEEs can cause various bond ruptures in DNA including sugar-phosphate (C-O),[13,20] sugar-nucleobase (C-N),[10,11,41] base-H,[42] etc. However, the bulk of the experimentally observed DNA-LEE interaction products are due to the C-N and the C-O bond cleavages.[13] Previous theoretical studies on gas phase (and under implicit solvation) purine nucleotides have concluded that a TNI centered on the nucleobase is formed upon LEE attachment.[27,43–45] The formation of the TNI is followed by the base release or the C-O bond rupture.[27] However, one alternate route can be the direct transfer of the electron from the doorway dipole bound state to the C-O or C-N $\sigma^*$-MO, leading to bond cleavage. The most obvious way of studying the bond cleavage mechanism is by constructing a potential energy curve as a function of the bond length.

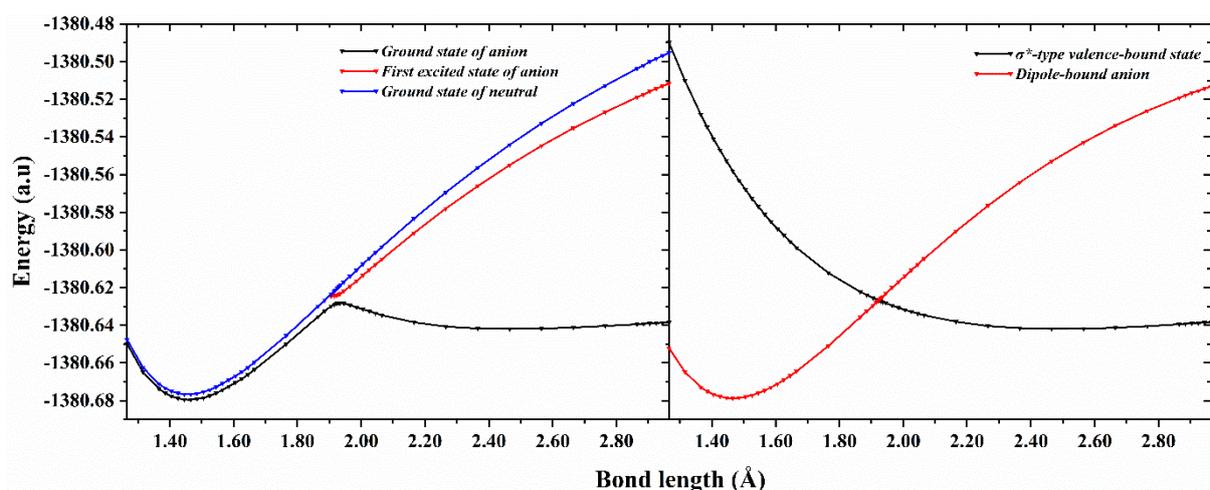

**Figure 3. Adiabatic (left) and diabatic (right) PEC for C-O bond dissociation from the dipole-bound state of 3′-dCMPH.**

Figure 3 represents the PEC for C-O elongation of 3′-dCMPH. The initial electron attached state is dipole-bound in nature. The stretching of the 3′C-O bond leads to the formation of $\sigma^*$ type valence-bound anionic state (See Figure 4), which becomes the ground state anion with the progressive increase in bond length. The adiabatic potential energy curve shows an avoided crossing between the dipole-bound and the σ* type valence-bound anionic states. Subsequently, the electron density gets transferred from the dipole-bound to the C-O σ* type valence-bound state because of the mixing of electronic and nuclear degrees of freedom. It leads to the rupture of the sugar-phosphate bond. While constructing the diabatic curve from the adiabatic picture, we had to choose the dipole-bound state and C-O σ* type valence-bound state as the diabatic basis. Morse potential as a function of the bond length was used to model the diagonal terms in the crossing model potential. The coupling constant has been found to be 45.59 meV, and the Marcus estimation of the rate of electron transfer is $9.04 \times 10^{-1} s^{-1}$. Therefore, the rate of electron transfer to C-O σ* leading to bond breaking is considered to be a much slower process than the formation of the stable anion.

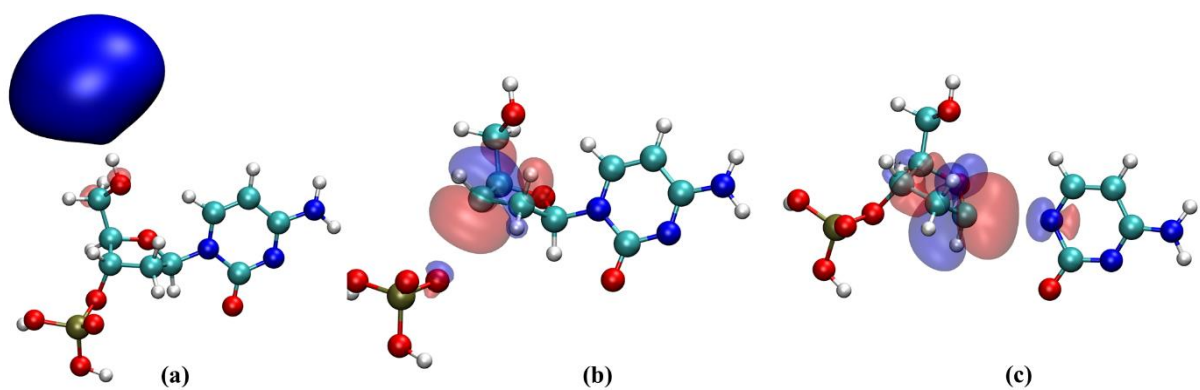

**Figure 4. The molecular orbital corresponding to the dominant transition in the EA-EOM-DLPNO-CCSD for the lowest energy anionic state (a) at dipole-bound anionic geometry, at (b) C-O bond length and (c) C-N bond lengths 1.5 Å longer than their neutral equilibrium value.**

We have observed a similar doorway mechanism for C-N bond breaking, where the electron transfer from the initial dipole bound state to the C-N σ* (See Figure 4(c)) leads to bond cleavage. Figure 5 shows the existence of an avoided crossing in the adiabatic potential energy curves corresponding to the C-N bond cleavage. We have performed a diabatization using the dipole bound and C-N σ* state as the diabatic basis, and the rate of electron transfer has been found to be $4.34 \times 10^{-14} s^{-1}$. Therefore, the rate of electron transfer leading to C-O bond cleavage in the doorway mechanism is higher than that in the C-N bond cleavage, which is consistent with the available experimental results.[13]

**Table 1. The coupling element and the rate constant values of C-O and C-N bond rupture in 5′-dCMPH, 3′-dCMPH, and NPN. NPN is the nucleoside-phosphate-nucleoside system.**

| Molecule | Bond | W (meV) | Rate Constant $\left(\text{s}^{-1}\right)$ |
|---|---|---|---|
| 5′-dCMPH | C-O | 8.92 | 2.95 |
|  | C-N | 19.11 | $1.80 \times 10^{-12}$ |
| 3′-dCMPH | C-O | 45.59 | $9.04 \times 10^{-1}$ |
|  | C-N | 35.07 | $4.34 \times 10^{-14}$ |
| NPN | 5′C-O | 15.04 | $2.91 \times 10^{-3}$ |
|  | 3′C-O | 37.36 | $7.97 \times 10^{-1}$ |

Similar trends were observed for the 5′-deoxycytidine monophosphate (5′-dCMPH), where the dipole-bound state acts as a doorway for electron capture, and the electron subsequently gets transferred to C-O or C-N $\sigma^*$ states which lead to bond breaking. From Figure S1 one can see that the adiabatic potential energy curves corresponding to C-O or C-N internuclear distance show avoided crossing, and the mixing of electronic and nuclear degrees of freedom leads to electron transfer. The rate of electron transfer to C-O $\sigma^*$ in 5′-dCMPH is much higher than that to C-N $\sigma^*$, similar to the trend observed in 3′-dCMPH.

The C-N bond rupture does not lead to SSB, but it can cause dangerous mutations in DNA. The C-O bond, however, is on the DNA-backbone and its cleavage would cause strand break. Moreover, there are two different C-O sites – at 3′ (3′C-O) and 5′ (5′C-O) positions of the sugar. Knowing which one is more prone to dissociation will help us to reconcile with the already existing understanding of DNA strand breaking. Earlier theoretical studies using a simple sugar-phosphate-sugar model with nucleobase replaced by -NH2 group indicated a resonance-based pathway for bond-cleavage, and the 5′C-O bond was reported to be more susceptible to cleavage.[46,47] However, Wagner and coworkers had experimentally shown that the 3′C-O bond is the favorable SSB site from the fragmentation pattern of thymidine nucleotide dimer[20], which contradicts the theoretically simulated results.

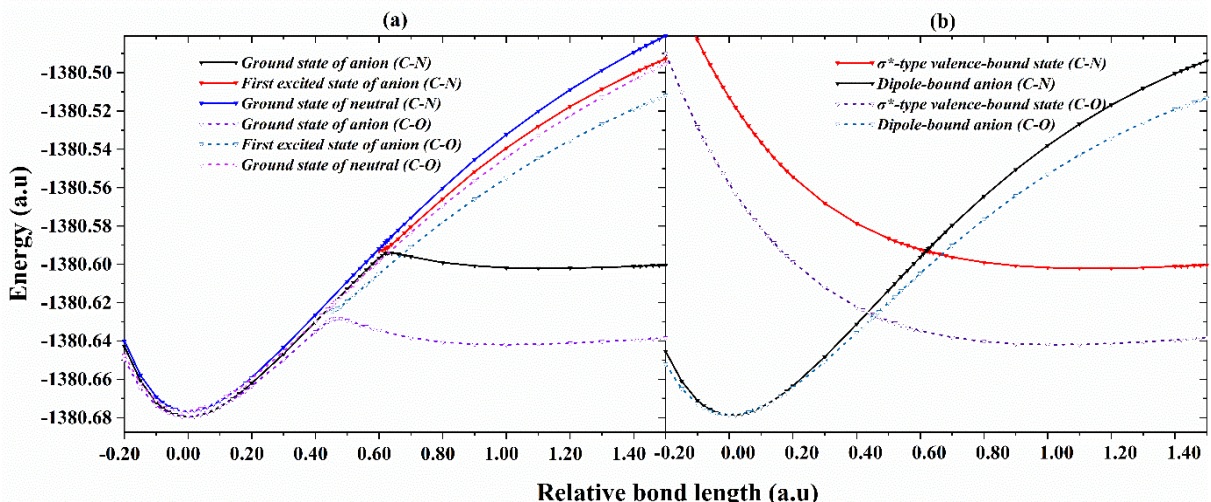

**Figure 5.** a) Adiabatic PEC and b) diabatic PEC for C-O and C-N bond cleavage for 3′-dCMPH. The bond length is shown relative to its equilibrium value.

We have chosen a more realistic nucleoside-phosphate-nucleoside (NPN) model system with cytidine as the nucleoside to determine the favorable SSB site between the two sugar-phosphate bonds. The electron attachment at the neutral geometry leads to the formation of a dipole-bound state (See Figure 6(A)). This dipole bound state acts as a doorway for electron capture, and the additional electron subsequently gets transferred to the 3′C-O $\sigma^*$ or 5′C-O $\sigma^*$, leading to bond cleavage. In Figure 7(a), we have plotted the adiabatic potential energy curve for the ground and the first excited states of the anion for the stretching of 3′C-O and 5′C-O bonds in the NPN model system. It shows avoided crossing between the ground and excited state of the anion similar to that observed in 3′-dCMPH and 5′-dCMPH.

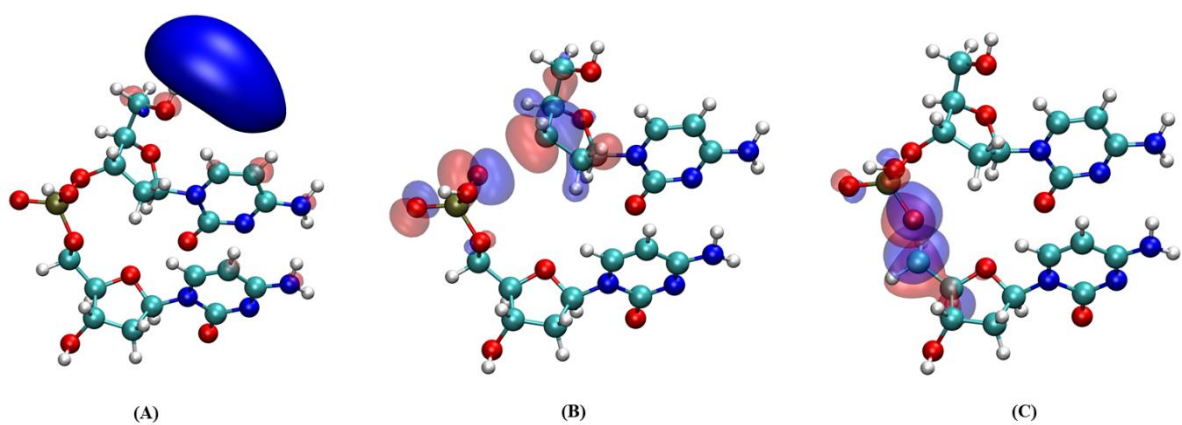

**Figure 6.** The molecular orbital corresponding to the dominant transition in the EA-EOM-DLPNO-CCSD for the lowest energy anionic state for NPN model system at (A) dipole-bound anionic geometry, at (B) 3′C-O and (C) 5′C-O bond lengths 1.5 Å longer than the equilibrium values.

We have employed a similar procedure mentioned earlier to model the diabatic PEC using the dipole-bound anion and the C-O $\sigma^*$ anionic state as the basis. The coupling constants calculated for the 3′C-O bond and 5′C-O bond are 37.36 meV and 15.04 meV, respectively. The rates of electron transfer to 3′C-O σ* is almost a hundred times higher than that to 5′C-O σ* state (See Table 1). Our findings are consistent with the experimental results of Wagner and coworkers[20] and contradict earlier theoretical predictions.[46,47]

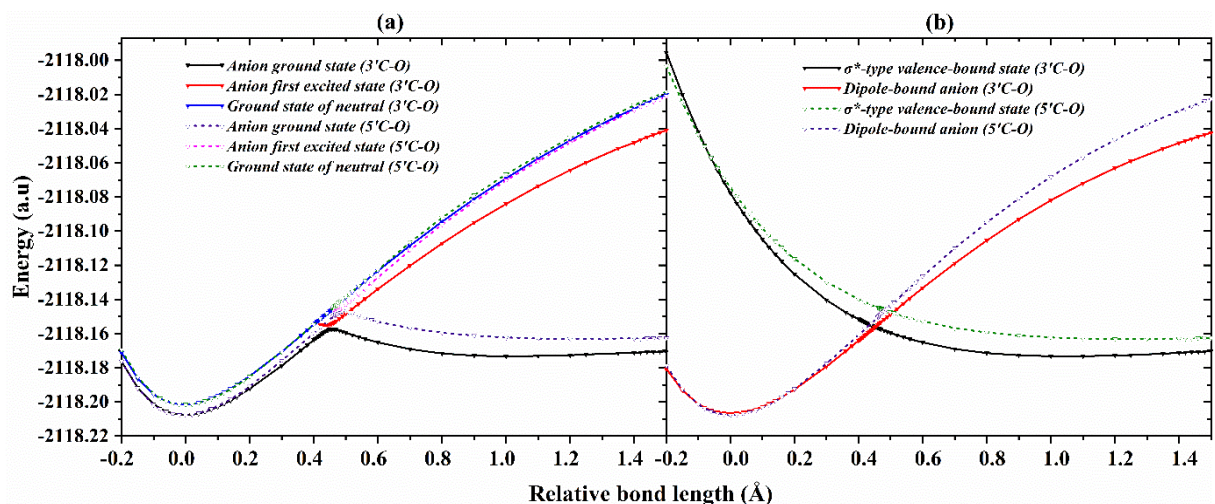

**Figure 7. a) Adiabatic and b) diabatic PEC for 5′C-O and 3′C-O bond cleavage for NPN system. The bond length is shown relative to its equilibrium value.**

For the cytosine nucleotide, the transition from dipole-bound to valence-bound anion is kinetically favored over the bond break at SSB sites. However, the formation of the π* type valence-bound anion does not eliminate the possibility of strand break because the additional electron might subsequently get transferred to a σ* state leading to C-O or C-N bond cleavage. While considering the kinetics of this process, one should expect this rate to be lower because high rates of SSB, if possible, would render the DNA vulnerable to radiation-induced DEA. In such a scenario, the DNA repair mechanism will be overwhelmed with the number of strand breaks and would certainly fail. We have considered a two-step process (Figure 8), where the valence-bound state is first formed from the doorway dipole-bound state. In the second step, the electron gets transferred from the π* valence-bound state to 3′C-O σ* orbital, leading to bond cleavage. The second step is a three-state process that involves the ground state, first excited state, and second excited state of the anion. Our estimated rate of electron transfer from the valence-bound state to σ* state leading to rupture of 3′C-O bond is $1.20 \times 10^{-13}$ s$^{-1}$. Thus, although the valence-bound 3′-dCMPH anion may also lead to 3′C-O bond cleavage in the gas-phase, the process is kinetically unfavorable. One should expect an even smaller rate for C-O bond dissociation in the condensed phase due to solvent stabilization of the anion.

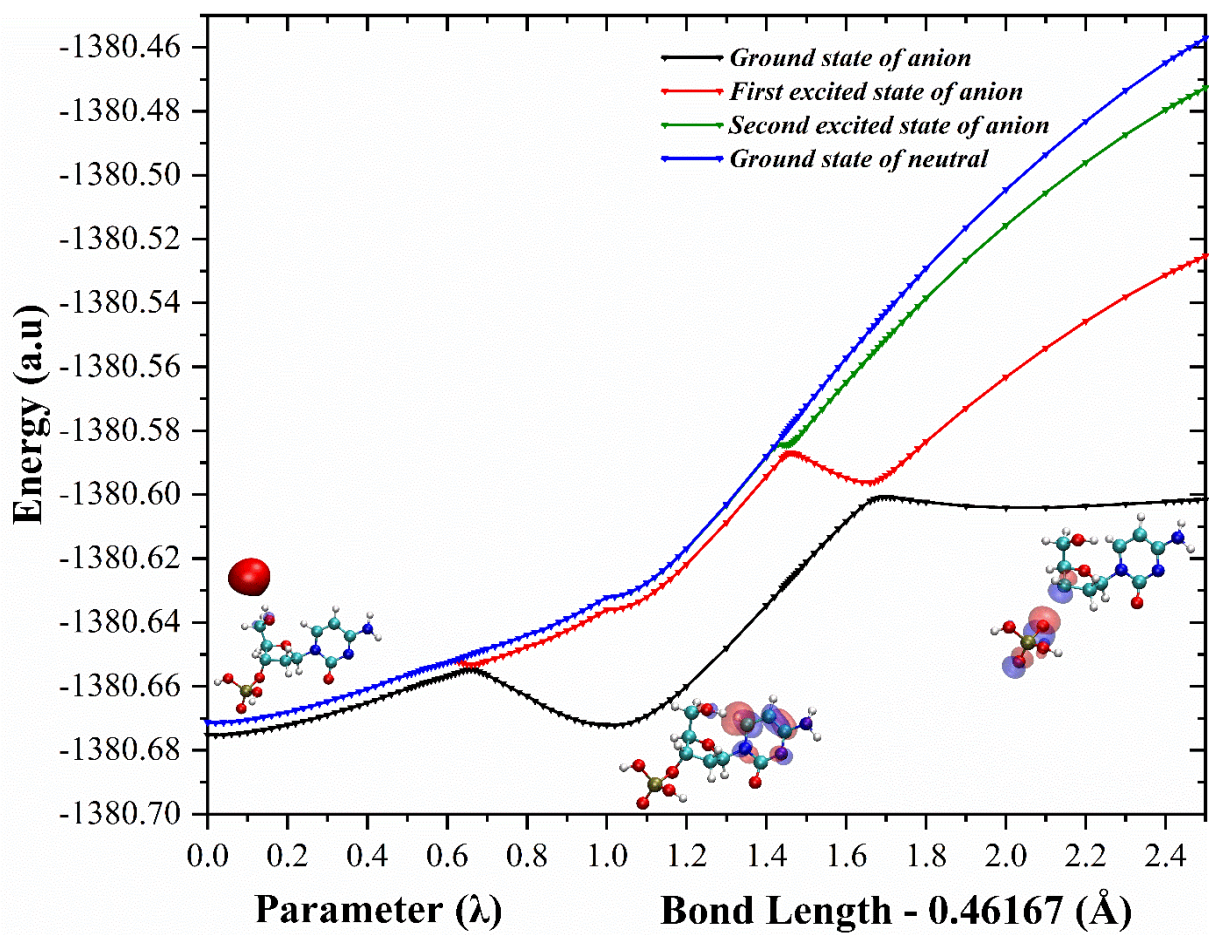

**Figure 8.** Adiabatic PEC showing the dipole-bound to valence-bound transition of 3′-dCMPH and the C-O bond dissociation from its valence-bound anionic state. The C-O bond length is scaled to make the graph continuous.

Our calculations show the existence of a doorway mechanism in low-energy electron attachment induced bond breaking in short DNA oligomers in the gas phase. The initial electron attachment leads to the formation of a dipole-bound state, which acts as a doorway for electron capture. Once the dipole-bound anion is formed, three competing processes may occur: (1) formation of a valence type π* anion, (2) formation of a C-O σ* type valence anion and subsequent C-O bond cleavage, and (3) formation of a C-N σ* type valence anion and subsequent C-N bond cleavage. The first process is kinetically favored over the other two, which can explain the resilience of the genetic material in our cells to DEA inducers. We have shown that in the doorway mechanism, the phosphodiester bonds are more likely to be cleaved than the sugar-nucleobase bond, and the 3′C-O breaks easily than 5′C-O. Both the observations are consistent with the available experimental results. The formation of valence π* type anion reduces the possibility of electron transfer to the dissociative σ* C-O/C-N state. Moreover, the stability of the valence π* type anion has been known to increase drastically in the aqueous phase,[36,37] which will enhance the rate of formation of stable π* type valence-bound anionic state in the solvent phase. This reduces the probability of C-O and C-N bond cleavage upon electron attachment in the aqueous phase and explains the suppression of SSB in the aqueous phase compared to the gas phase.[28,48] The doorway mechanism described above will be competitive to the already known resonance-based pathway of SSB. The consideration of both pathways is essential to get a complete understanding of the low-energy electron-induced SSB. It should be noted that the rate constant calculated using rigid potential energy surface scans in conjugation with Marcus theory is rather qualitative. One needs to perform non-adiabatic molecular dynamic simulations, including the effect of the surrounding environment, to get a quantitative description of the doorway mechanism for SSB. Work is in progress towards that direction.

**Computational Details**

The negative charge of 3′-dCMP and 5′-dCMP were neutralized by protonating the oxygen of the phosphate group (denoted as 5′-dCMPH and 3′-dCMPH). The neutral and anion geometries of both molecules were optimized at RI-MP2/def2-TZVP level of theory. A subsequent frequency calculation showed no imaginary frequencies, which ensured that the optimized geometries are at local minima. A rigid potential energy surface scan of the sugar-phosphate (C-O) and sugar-nucleobase (C-N) bond cleavage was carried out for both 3′-dCMPH and 5′-dCMPH. C-O and C-N bonds were increased in steps of 0.1 Å to generate the intermediate geometries. The bond length in the final geometry was 1.5 Å longer than the equilibrium bond length. The vertical electron affinities were calculated using EA-EOM-DLPNO-CCSD level of theory[49] with the aug-cc-pVDZ basis set. An additional 5s5p4d functions were added to the atom closest to the positive end of the molecular dipole moment vector. We have used a development version of ORCA[50,51] to perform the calculations. The intermediate geometries in the linear transit between the dipole-bound and valence-bound anions are obtained using the following expression

$$R = (1-\lambda)R_{DB} + \lambda R_{VB} \qquad (1)$$

Here, $R$ represents a parameter of the intermediate geometry. It could be the bond length, bond angle, or dihedral angle. $R_{DB}$ and $R_{VB}$ are the value of geometric parameters in the dipole-bound and valence-bound geometries, respectively. $\lambda$ is varied from zero to one, to obtain the intermediate geometries. The $\lambda = 0$ corresponds to dipole-bound geometry, and the valence-bound anion is obtained when the $\lambda$ equals 1. The coupling between the dipole and valence-bound state has been calculated using a simple two state avoided crossing model:

$$V = \begin{pmatrix} V_1 & W \\ W & V_2 \end{pmatrix}$$

Here, $W$ is the coupling element. $V_1$ and $V_2$ are diagonal terms calculated using the Morse potential expression as a function of $\lambda$:

$$V_i = \omega_i\left(1 - e^{-a(\lambda - \lambda_i^0)^2}\right) + v_i^0 \qquad (2)$$

We have calculated the approximate rate of dipole-bound to valence-bound transition and bond cleavage using Marcus theory as described in ref [34]. The expression for rate constant, $k$ is as follows:

$$k = \frac{2\pi}{\hbar}|W|^2 \sqrt{\frac{1}{4\pi k_B T \lambda_R}} e^{-(\lambda_R + \Delta G^0)^2/4\lambda_R k_B T} \qquad (3)$$

where $\lambda_R$ is the reorganization energy, and $\Delta G^0$ is the free energy change between the valence-bound and dipole-bound states $(E_{VB} - E_{DB} - \Delta S)$. $\Delta S$ is the entropy contribution which was neglected in the case of dipole-bound to valence-bound transition of 3′-dCMPH, as it is generally found to be small for cases where the structural changes only involves distortion of a ring. For bond cleavage process, $\Delta S$ was calculated at the level of theory used for optimization.

**Supporting Information**

The Supporting Information is available.

- Details of the procedure used for the construction of the NPN model system.
- Cartesian coordinates of the optimized neutral and anion geometry of 3′-dCMPH, and neutral geometries of 5′-dCMPH, NPN model system.
- Adiabatic and diabatic PEC of C-O and C-N bond cleavage in 5′-dCMPH.

**Acknowledgments**

The authors acknowledge the support from the IIT Bombay, IIT Bombay Seed Grant project, DST-SERB, DST-Inspire Faculty Fellowship for financial support, IIT Bombay super computational facility, and C-DAC supercomputing resources (PARAM Yuva-II, PARAM Brahma) for computational time.

**Conflict of interest**

The authors declare no competing financial interest.